\documentstyle[10pt]{article} 

\begin{document} 

\begin{center} 

       {\bf   The two types of Cherenkov gluons at LHC energies}    \\

                           I.M. Dremin\\

              {\it  Lebedev Physical Institute, Moscow \\
              dremin@td.lpi.ru}

\end{center} 

\begin{abstract} 
Beside comparatively low energy Cherenkov gluons observed at RHIC, there could 
be high energy gluons at LHC, related to the high energy region of positive 
real part of the forward scattering amplitude. In both cases they give rise
to particles emitted along some cone. The characteristics of the cones 
produced by these two types of gluons are different. Therefore different 
experiments are needed to detect them. The cosmic ray event which initiated 
this idea is described in detail.
\end{abstract} 

RHIC experiments \cite{5, 6} have shown the two-bump structure of the 
azimuthal angle distribution near the away-side jets. This can be interpreted 
as the emission of Cherenkov gluons \cite{3,4,a,7}. The notion of Cherenkov 
gluons was first introduced in \cite{1, 2} to explain some cosmic 
ray data \cite{8}. 

Analogous to Cherenkov photons, the Cherenkov gluons can be emitted in 
hadronic collisions provided the nuclear index of refraction $n$ exceeds 1. 
The partons moving in such nuclear medium would emit them. 
These gluons should be emitted at the cone surface with the cone angle 
$\theta $ in the {\bf rest system} of the {\bf infinite} medium defined by 
the relation 
\begin{equation} 
\cos \theta = \frac {1}{\beta n},   \label{cos} 
\end{equation} 
where $\beta $ is the ratio of the velocities of the parton-emitter and 
light which can be replaced by 1 for relativistic partons. 

The gluon propagation in a nuclear medium can be described by the nuclear index
of refraction. It is well known \cite{gwa} that the  excess $\Delta n_R(E)$
of $n$ over 1 is proportional to the real part of the forward scattering 
amplitude ${\rm Re} F(E)$ and to the number of scatterers $\nu $. It must be
positive for Cherenkov gluons to be emitted:
\begin{equation} 
\Delta n_R(E)=n-1\propto \nu (E){\rm Re} F(E)>0.  \label{DN} 
\end{equation} 
Here we must use the forward scattering amplitude of a gluon propagating in 
strong fields of the quark-gluon medium formed in heavy-ion collision.
QCD estimates of $F(E)$ are rather indefinite. Therefore, we have to rely on 
general properties of hadronic reactions known from experiment. According to 
our knowledge about hadronic reactions, ${\rm Re} F(E)$ is positive for any 
colliding hadrons either within the lower wings of the Breit-Wigner 
resonances \footnote{At the maximum of the resonance ${\rm Re } F(E_R)=0$ as 
seen below in Eq. (\ref{BW}). In electrodynamics this is used, in particular, 
to solve the problem of abundance of elements in the Universe \cite{fey}.} or 
at very high energies \cite{dnaz, bloc}. Jet energies available in RHIC 
experiments \cite{5,6} are sufficient only for emission of gluons with rather 
low energy. The resonance production is the common property of hadronic 
interactions at low energies. Thus one can hope that the same feature is valid 
for gluons as the carriers of strong forces. Therefore, we attribute the above 
mentioned specifics of RHIC data to resonance effects as discussed in \cite{3}. 
 
The energies of colliding partons at LHC are much higher. They are sufficient
for emission of high energy Cherenkov gluons. Therefore, one could await for 
two types of Cherenkov gluons produced corresponding to these low and high 
energy regions. At LHC these effects can be observed, correspondingly, in 
trigger and inclusive experiments as argued below. Namely this idea about 
high energy Cherenkov gluons was promoted for explanation of peculiar features 
of the ring-like particle distribution in a cosmic ray event with energy 
close to those of LHC \cite{1}.

Let us explain it again in more detail here. In distinction to "in-vacuum"
jets described by DGLAP equations, the collective effects of gluon emission
can be pronounced in a medium. An initial parton of the impinging nucleus
enters the medium (another nucleus) and emits gluons with some energy spectrum.
According to Eq. (\ref{DN}) the nuclear index of refraction depends on energy.
In those regions of the spectrum, where this index of refraction $n$ exceeds 1, 
the Cherenkov mechanism of gluon emission can be at work. Since QGP is considered 
as a medium filled mostly by gluons one can say that the
energies of emitted Cherenkov gluons should belong to those regions where 
the real part of the forward scattering amplitude of the Cherenkov gluon and 
its partner in the medium is positive. E.g., this happens if the two gluons
(subject to strong but soft surrounding fields!) have relative energy 
which fits the $\rho $-meson's left wing. The collective properties of gluons 
in sQGP determine that these Cherenkov gluons should be emitted at a definite 
polar angle and hadronize forming $\rho $-mesons by their convolution with a 
"gluon" of the medium.

We do not require $\rho $-mesons or other resonances 
pre-exist in the medium but imply that they are the modes of its excitation 
formed due to the hadronization properties of partons. 
The Cherenkov gluon emission is a collective response of the quark-gluon medium 
to impinging partons related to its hadronization. 

According to Eq. (\ref{cos}) the ring-like 
distribution of particles must be observed in the plane perpendicular to 
the momentum of the parton-emitter. If projected on the ring diameter, it 
results in the two-bump structure observed at RHIC. In more detail it
has been discussed in \cite{3,4,a,7}. Some predictions and proposals are
put forward in \cite{sh}. Here I describe briefly the conclusions drawn
from RHIC experiments and then concentrate on LHC specifics.

RHIC experiments were done with a trigger placed at 90$^o$ to the collision 
axis. It implies that initial colliding partons suffered hard scattering at 
large angles. Its probability is low but the background is strongly diminished 
as well. The emission of Cherenkov gluons as the collective response of the 
medium to propagation of the scattered away-side parton should give rise to 
the ring-like
distribution of particles in the plane perpendicular to the away-side jet axis.
This plane is orthogonal both to the collision plane (formed by the collision 
and trigger jet axes) and to the azimuthal plane. Therefore the ring projection
on azimuthal plane used in RHIC experiments which resulted in two peaks in
azimuthal angles is not at all the optimum one. It would be better to use the 
away-side jet direction as $z$-axis and plot the traditional pseudorapidity
distribution. RHIC data show the distance between the peaks defined in angular 
variables ($\theta =D$ in PHENIX notations). Herefrom one gets according to 
Eq. (\ref{cos}) the nuclear index of refraction. Its value is found to be 
quite large $n=3$ compared to usual electromagnetic values close to 1
for gases. 

Nevertheless, for preliminary estimates let us use the traditional expressions
\cite{gwa, fey} of electrodynamics for dilute media. RHIC experiments deal 
with rather soft gluons because jet energies are typically about several GeV 
only. Therefore one can replace ${\rm Re} F(E)$ in (\ref{DN}) by the sum
over the Breit-Wigner resonances (see \cite{3, 7}):
\begin{equation} 
{\rm Re} n(E)=1+\sum _R\frac {2J_R+1}{(2s_1^R+1)(2s_2^R+1)}\frac {6m_{\pi }^3
\Gamma _R \nu } {EE_R^2}\frac {E_R-E}{(E-E_R)^2+\Gamma _R^2/4}. \label{BW} 
\end{equation} 
Here $E_R$, $\Gamma _R$ are the position and the width of the resonance $R$, 
$J_R,\, s_1^R,\, s_2^R$ are spins of the resonance and its decay products.

The rather high value of $n$ results in the large density of partons in the 
created quark-gluon system with $\nu \approx $20 partons within the volume 
of a single nucleon. The overwhelming contribution in this estimate 
according to (\ref{BW}) is given by the $\rho $-meson.

This value of $\nu $ really shows that one has to deal not with a gas but
with a liquid-like matter.
For dense media $\Delta n_R$ increases faster than linearly with $\nu $.
Since $n$ is large the color polarizability $\alpha $ must play an important 
role in sQGP. One should use the analogue of the electrodynamic formula
\cite{gwa, fey}:
\begin{equation}
\frac {n^2-1}{n^2+2}=\frac {m_{\pi }^3\nu \alpha }{4\pi }=\sum _R
\frac {2J_R+1}{(2s_1^R+1)(2s_2^R+1)}\cdot \frac {4m_{\pi }^3\Gamma _R \nu }
{EE_R^2}\cdot \frac {E_R-E}{(E-E_R)^2+\Gamma _R^2/4}.   \label{liq}
\end{equation}
From it one gets almost twice smaller value of $\nu $ (about 10 partons per 
nucleon). Still our conclusion about the liquid-like medium is valid but
this liquid is not extremely dense.

This value of $n$ is also related to the energy loss of gluons 
estimated in \cite{7} as 
\begin{equation}
\frac {dE}{dx}=4\pi \alpha _S\int _{E_R-\Gamma _R}^{E_R}E(1-\frac {1}{n^2})dE\approx 1 GeV/Fm.     \label{dedx}
\end{equation}
The height of the peaks 
determines the width of the ring. In its turn, it defines the free path 
length of Cherenkov gluons \cite{7} which is long enough 
$R_f\sim 7$ Fm. Thus one obtains knowledge about such crucial properties of 
the hadronic medium as its index of nuclear refraction, density of partons, 
free path length and energy loss. The unusual particle content within the 
ring with abundant production of shifted to lower masses and squeezed resonances 
(mostly $\rho $-satellites at about 690 MeV) is also predicted. This specific 
feature can be checked by studying the mass distribution of $e^+e^-$-pairs in 
this particular mode of resonance decay (see \cite{sh}).

The estimated energy loss is not very high and the free path length is
large enough for gluons to hadronize near the surface of the QGP volume.
The imaginary part of $n$ is about three times less than its real part
if the formula (\ref{liq}) and corresponding expression for ${\rm Im } n(E)$
are used: 
\begin{equation} 
{\rm Im} n(E)=\frac {2J+1}{(2s_1+1)(2s_2+1)}\frac {3m_{\pi }^3 
\nu } {EE_R^2}\frac {\Gamma_R^2}{(E-E_R)^2+\Gamma _R^2/4}. \label{im} 
\end{equation} 
This statement can sound strange if compared to the footnote
comment. However the main difference is provided by the shift of the maximum
for ${\rm Re } n(E)$ relative to the usual resonance mass and diminished role
of the imaginary part at this maximum. This favors smaller absorption.
That is why we attribute the effect of Cherenkov gluons to hadronization and
collective properties of sQGP.

The similar experiments with trigger jets at 90$^o$ to the collision axis 
are possible at LHC. Their results must be the same as RHIC data even for 
higher energy jets if the parton density does not change with energy. The 
effect is saturated by low energy resonances and does not depend on jet 
energy. 
These trigger experiments require high luminosity because the two jets are 
produced by the large angle scattering of initial partons with small cross 
section. The medium is at rest on the average in the c.m.s.. Influence of its 
internal flows on the cone angle can be accounted \cite{ssm}. 

There is no luminosity problem for forward (and backward) moving initial partons. 
They are much more abundant and able to produce both low and high energy 
Cherenkov gluons. The background is much larger, however.
The medium rest system where Eq. (\ref{cos}) is applicable 
now is the rest system of one of the colliding partners. Namely in this system 
the angle of emission of low energy Cherenkov gluons should be about 70$^o$ 
(for $n=3$) to the collision axis. When transformed to the c.m.s., 
it is in the deep fragmentation region of pseudorapidity above 8 which is 
inaccessible for any LHC detector. Thus the large angle trigger experiment 
is the only way to observe low energy Cherenkov gluons at LHC. The dependence 
of the Cherenkov cone angle on the angle of the trigger jet would be
interesting to measure. It would reveal the motion of the medium as a whole. 

The ingenious approach with large angle parton scattering at RHIC was necessary
to detect the low energy gluons and diminish the background. At LHC, one can 
also try to detect Cherenkov effect in inclusive experiments. Any parton moving
in the nuclear medium can emit gluons. The partons moving along the collision 
axis are most abundant and possess highest energies. At LHC they can emit 
gluons with very high energy. For them the nuclear index of refraction can also 
exceed 1 \cite{1, 2}. This statement is based on the experimentally found 
general property of all hadronic reactions supported by the dispersion 
relations \cite{dnaz, bloc} that the real part of the forward scattering 
amplitude is positive at very high energies. Thus one can await for the high 
energy Cherenkov gluons emitted. Each of them would produce a 
jet (in place of a resonance for low energy gluons). The specific feature 
of these jets would be their tendency to lie on a cone, i.e. to form the ring 
filled in by "spots" (dense groups of particles) in the plane perpendicular to 
the collision axis. The ring parameters differ from those for low energy gluons. 

The high energy behavior of $\Delta n_R$ was 
estimated in \cite{1,2} purely phenomenologically using various experimental 
data on hadronic reactions and dispersion relation predictions. The properties
of a gluon jet traversing the nuclear medium were identified with properties of
hadrons, scattering on hadronic targets. Then the dependence of the nuclear 
index of refraction on jet energy $E_j$ in the target rest system can be
approximated as
\begin{equation}
\Delta n_R(E_j)\approx \frac {a\nu_h}{E_j}\theta (E_j-E_{th})   \label{nh}
\end{equation}
above some threshold energy $E_{th}$ about 70 - 300 GeV. It implies 
that only jets with energies exceeding this threshold can be 
produced according to Cherenkov effect. Here, $a\approx 2\cdot 10^{-3}$ GeV 
is a fitted parameter with $\rho (E)\approx 0.1$
and $\nu _h$ is the parton density for high energy region. It can differ
from $\nu $ used in the low energy region. However these estimates are rather 
indefinite and it is better to rely on the analysis of the cosmic ray event. 

The idea about Cherenkov gluons was used to interpret the cosmic ray event 
at energy about 10$^{16}$ eV \cite{8} where two rings more densely populated by 
particles than their surroundings were noticed. This event was registered in 
the detector with 
nuclear and X-ray emulsions during the balloon flight at the altitude about
30 km. The most indefinite characteristics of the event is the height $H$
over the detector at which the interaction took place. However, it can be
estimated if one assumes that the two rings observed with radii $r_1=1.75$ cm
and $r_2=5$ cm are produced by forward and backward moving (in c.m.s.) partons, 
correspondingly. Using the transformation of angles from target ($t$) to 
c.m.s. ($c$) 
\begin{equation}
\tan \frac {\theta _c}{2}\approx \gamma \theta _t   \label{tan}
\end{equation}
and assuming the symmetry of rings $\theta _{2c}=\pi - \theta _{1c}$, one gets 
$\theta _{1c}\approx 61^o$ and $\gamma \approx 2.3\cdot 10^3$.
The target rest system angle is $\theta _{1t} \approx 2.6\cdot 10^{-4}$ 
and the height $H=r_1/\theta _{1t} \approx 68$ m. 
Both the energy and the height correspond rather 
well to the experimental estimates \cite{8} recently revised, confirmed and 
obtained by three different methods \cite{kon}. The most reliable of them give 
height values ranging from 50 m to 100 m. The medium rest system coincides 
with the target rest system, and the values of emission angles indicate on 
smaller (than $\nu $) density of scatterers if (\ref{nh}) is used.
 
Even though the observed angles were quite small in the target rest system, 
at LHC they would correspond to large c.m.s. angles about 60$^o$ - 70$^o$. 
The background at LHC was estimated as rather flat according to HIJING model 
\cite{9}. The peaks in the angular (pseudorapidity) distribution of jets over 
the background at these angles would be observable. The ring-like and spotty
two-dimensional plot leading to such maxima was confirmed 
by the wavelet analysis of the cosmic ray event \cite{adk}. Let us note that 
some indications on similar effects in hadronic reactions at lower energies 
were obtained with a special selection of events in the high statistics sample 
\cite{ag} as well as by the wavelet analysis of individual events \cite{dkp}. 

In connection with Cherenkov gluons we discussed only those energy regions
where ${\rm Re } F(E)>0$. However there exists the transition radiation which 
can be emitted also in the regions of the negative values of the real part
of forward scattering amplitude. It was argued in \cite{2} that this effect
is negligibly small because it is proportional to $\Delta n_R^2$ for small 
$\Delta n_R$, i.e. for high energy gluons. However we have found that at
rather low energies $n(E)$ can substantionally exceed 1. Thus the transition 
radiation can become also important. This problem has not been solved.

To conclude, we claim that low energy Cherenkov gluons will be observed at LHC 
in the trigger experiments with trigger jets emitted at large angles to the 
collision axis analogously to RHIC experiments. Their properties would indicate 
how the nuclear index of refraction and, consequently, the density  of 
scattering centers (gluons, quarks, vacuum fields ...?) and
other characteristics of the nuclear medium change with energy if compared 
with RHIC data. The dependence of the cone angle on the trigger jet angle 
would show the effects of the medium motion. 

The non-trigger inclusive experiments can reveal the high energy Cherenkov 
gluons at LHC by observing the excess of jets at fixed and comparatively 
large angles in the c.m.s. Such gluons are unobservable at RHIC because 
of the energy threshold. The energy dependence of main properties of the
nuclear medium at very high energies will be demonstrated at LHC.

This work was supported in part by grants RFBR 05-02-39028-GFEN, 06-02-17051 
and NNSF of China under project 10475030.

\end{document}